\documentclass[amsmath,amssymb,preprint]{revtex4-1}

\usepackage{graphicx}
\usepackage{dcolumn}
\usepackage{bm}
\usepackage{subfigure}
\usepackage{empheq}
\usepackage{url}
\usepackage[utf8]{inputenc}
\usepackage[T1]{fontenc}
\usepackage{mathptmx}
\usepackage{xcolor}

\begin{document}
\title[]{Control of cascading failures in dynamical models of power grids}
\author{M. Frasca}\email{mattia.frasca@dieei.unict.it}
\affiliation{Dipartimento di Ingegneria Elettrica, Elettronica e Informatica, Universit\` a degli Studi di Catania, Catania, 95125, Italy}
\affiliation{Istituto di Analisi dei Sistemi ed Informatica ``A. Ruberti'', Consiglio Nazionale delle Ricerche (IASI-CNR), 00185 Roma, Italy}

\author{L. V. Gambuzza}\email{lucia.gambuzza@dieei.unict.it}
\affiliation{Dipartimento di Ingegneria Elettrica, Elettronica e Informatica, Universit\` a degli Studi di Catania, Catania, 95125, Italy}

\begin{abstract}
In this paper, we introduce a distributed control strategy to prevent dynamically-induced cascading failures in power grids. We model power grids using complex networks and nonlinear dynamics to provide a coarse-grained description of the electro-mechanical phenomena taking place on them (in particular, we use coupled swing equations) and restrict our analysis to cascades of line failures, i.e., failures due to power flows exceeding the maximum capacity of a line. We formulate a distributed control protocol relying on the same topology of the physical layer and apply it to several power grid models, including a small-size illustrative example with five nodes, the Italian high-voltage (380kV) power grid, and the IEEE 118-bus system. Our results indicate that the approach is capable of preventing cascading failures, either controlling each node of the network or a suitable subset of them. 
\end{abstract}

\maketitle

\section{Introduction}

A power grid is a system of coupled electrical devices that deliver power from generators to consumers. It can be represented in terms of a networked system, in which the nodes are points where the power is injected by a generator or extracted by consumers, and links represent the electrical connections between pairs of nodes, modeling transmission lines \cite{nishikawa2015comparative}.
As the normal functioning of a power grid corresponds to synchronization of all nodes, an important characteristic of the power grid is the ability to reach and maintain the synchronous state, even in the presence of different types of disturbances that may affect it. The way in which the dynamics of a power grid is described is, however, not unique, and several models have been considered, mainly differing for the modeling approach adopted for the loads \cite{nishikawa2015comparative}.

Many works on synchronization on power grids \cite{dorfler2012synchronization,grzybowski2016synchronization,gambuzza2017analysis,schafer2018dynamically,totz2020control} adopt a dynamical description of the power system based on a network of (second-order) Kuramoto oscillators with inertia, which derives from the swing equation governing the rotor mechanical dynamics
\cite{filatrella2008analysis}. This model allows for the study of the conditions and the role of the parameters of the dynamical units under which the power network remains synchronized. For instance, in \cite{dorfler2012synchronization} sufficient algebraic conditions leading to synchronization are introduced, while transient stability analysis is obtained via singular perturbation analysis. In \cite{motter2013spontaneous}, instead, the conditions on tunable parameters of the dynamical units for spontaneous synchrony in power grids are investigated, showing that the stability of the synchronous state may be enhanced specifically by tuning two parameters of the generators. The role of the coupling strength in the onset of synchronization and the characteristics of network and oscillators favoring synchronization are studied in \cite{grzybowski2016synchronization}. Furthermore, in \cite{rohden2012self} stability against perturbations and robustness against structural damages are dealt with, finding that more decentralized grids are more robust to topological failures, but, at the same time, more sensitive to dynamical perturbations \cite{rohden2012self}. A different modeling approach is, instead, adopted by Lozano and coauthors \cite{lozano2012role} that establish the conditions for which a first-order Kuramoto network may be used to approximate the dynamics of the power system for the study of synchronization. With their model they characterize robustness of synchronization after removal of some lines, showing how, depending on the specific features of the links removed, this can lead either to an increase or a decrease of the synchronization threshold.

The problem of cascading failures, i.e. a sequence of failures and consequent disconnections of the networks after an initial outage, is particularly relevant in power grids and, as such, has been investigated in many works. However, in many of them the dynamics of the power system is often neglected. Several works, for instance, concentrate on the structural properties of the network, modeling a failure as the removal of a component (either a node or a link) of the network and, then, calculating its impact on the load redistribution \cite{motter2004cascade,crucitti2004model,dobson2007complex,zhang2016optimizing,buldyrev2010catastrophic,pahwa2014abruptness,witthaut2015nonlocal}. 
Other works incorporate a quasi-static description of the electrical properties of the system, considering the power flows obtained by solving the DC or AC power flow equations \cite{hines2010topological,soltan2015analysis,rohden2016cascading,cetinay2017comparing,cetinay2018nodal,strake2019non}. Studies that explicitly take into account the dynamics of the electro-mechanical phenomena occurring in the power grid adopt different levels of description, including very detailed characterizations of the circuitry involved in the power grid \cite{song2015dynamic} and more abstract models providing a simplified coarse-grained description of the power grid dynamics \cite{demarco2001phase,schafer2018dynamically}. 


The design of control strategies to prevent cascading failures is strictly connected to the model adopted for their description. An early example is discussed in \cite{motter2004cascade}, where the solution for cascade prevention, based on a selective, intentional removal of nodes and edges from the network after the initial failure, is tailored on the mechanisms unveiled by the analysis of the cascading phenomenon. A more recent example is provided in \cite{chen2020nonlinear}, where the conditions for stopping the spread of a failure are derived from the analysis of the nonlinear weighted network model explicitly incorporating the redundant capacity for the edges. 

In this paper, we investigate the problem of controlling dynamically-induced cascading failures in a power grid model. These are failures induced by an initial fault in one of the grid lines (e.g., due to some exogenous event) that then can lead to subsequent failures of other lines as a consequence of load redistribution. To tackle this problem, we consider the model discussed in \cite{schafer2018dynamically} where the units of the system are described by synchronous machines and line failures by an overload condition. This model clearly constitutes a simplification both of the power grid dynamics and the types of failures that may take place in it, for instance, in neglecting the presence of several protection and control mechanisms operating in real power systems or the different dynamics along the units. This simplified model still provides a coarse-grained description of the cascading failure phenomenon in a formalism that can be dealt with the tools of nonlinear dynamics and complex network theory. Here, we propose a solution based on the introduction of a second layer of distributed controllers, acting in parallel with the physical one. This scheme leverages the theoretical framework of multi-layer control for complex networks introduced in~\cite{lombana2014distributed} and applied in~\cite{totz2020control} to control synchronization in power grids. In particular, our control layer provides an input based on the actual frequency differences with neighboring nodes, thus reinforcing the tendency of the network to synchronization while reaching the new equilibrium state.




\section{Model of the power grid}
\label{sec:modelpowergrid}

To model the power grid, we consider the so-called synchronous machine model \cite{nishikawa2015comparative}, which has been proven to be particularly effective in the study of cascading failures \cite{schafer2018dynamically}. This model, in fact, crucially incorporates the node dynamics accounting for the time evolution of the voltage phase angles, that plays a fundamental role in the emergence of cascading failures.

In this model of the power grid, each node is associated with a rotating machine whose dynamics is described by a swing equation. In more detail, let $\mathcal{N}$ indicate the set of nodes with cardinality $|\mathcal{N}|=N$, and let $\mathcal{N}_g$ (with $|\mathcal{N}_g|=N_g$) represent the subset of generator nodes. Each node is characterized by a mechanical rotor angle that corresponds to the voltage phase angle and is indicated as $\theta_i(t)$, and by its angular velocity $\omega_i=d\theta_i/dt$, with $i=1,\ldots,N$, relative to a rotating reference frame with velocity $\Omega=2\pi f$, where $f=50Hz$ or $f=60Hz$ depending on the geographical area under study. Each rotating machine is also characterized by its inertia $I_i$, the damping coefficient $\gamma_i$, and the power $P_i$, which can be either positive, $P_i>0$, for nodes that act as generators, injecting the power into the system, or negative, $P_i<0$, for nodes that act as loads, absorbing the power from the system. The dynamics of the variables at each node are described by the following swing equation \cite{filatrella2008analysis}:

\begin{equation}
\label{eq:swingequations}
\begin{array}{l}
\frac{d\theta_i}{dt}=\omega_i\\
I_i\frac{d \omega_i}{dt}=P_i-\gamma_i\omega_i+\sum\limits_{(i,j)\in \mathcal{E}} K_{ij}\sin(\theta_j-\theta_i)
\end{array}
\end{equation}

\noindent where $\mathcal{E}$ represents the set of the operating links of the power grid and $K_{ij}$ are the elements of a weighted adjacency matrix describing its topology. The terms $K_{ij}$ are related to the electrical quantities characterizing the nodes by the relationship  $K_{ij}=B_{ij}V_iV_j$ where $B_{ij}$ is the susceptance between nodes $i$ and $j$, and $V_i$ and $V_j$ are the voltage amplitudes at the nodes. Eqs. (\ref{eq:swingequations}) apply under specific assumptions that simplify the  general power flows equations governing the electrical system. In particular, the voltage amplitudes $V_i$ are assumed to be constant, the ohmic losses negligible, and the variations in the angular velocities, $\omega_i$, small compared to the reference $\Omega$.

From the values of the angles at each time instant, the flow on the line $(i,j)$ is calculated as:
 
\begin{equation}
\label{eq:flowdefinition}
F_{ij}(t)=K_{ij}\sin(\theta_j(t)-\theta_i(t))
\end{equation}

The flow represents an important quantity to study failures in the power grid. From (\ref{eq:flowdefinition}), the maximum flow that a line can accommodate is given by $F_{ij}=K_{ij}$. However, since ohmic losses induce overheating in the line, the connection is shutdown when the flow exceeds a fraction $\alpha \in [0,1]$ of its maximum, which corresponds to set the line capacity as $C_{ij}=\alpha K_{ij}$. Here, $\alpha$ represents a tunable parameter of the model. The overload condition of a generic line $(i,j)$ can be, therefore, expressed as:

\begin{equation}
\label{eq:overloadcondition}
|F_{ij}(t)|>C_{ij}=\alpha K_{ij}
\end{equation}


The desired working operation for the power grid is characterized by all the units oscillating at the same frequency. As the system variables are represented by quantities relative to the reference angular velocity $\Omega$, this condition corresponds to a fixed point in the swing equations with $\omega_i=0$ and constant phases that can be determined by solving the following equation:
\begin{equation}
\label{eq:DCequilibrium}
0=P_i+\sum\limits_{(i,j)\in \mathcal{E}} K_{ij}\sin(\theta_j^*-\theta_i^*) 
\end{equation}

\section{Problem formulation}
\label{sec:problemFormulation}

In the absence of any control, a fault in a line of the synchronous-machine model~(\ref{eq:swingequations}) can produce a change in the angles such that the overload condition~(\ref{eq:overloadcondition}) is met at some $t$ for some line $(i,j)$ of the network; this in turns may trigger the failure of other lines, leading to a cascade of faults~\cite{schafer2018dynamically}. A crucial aspect of the model is that, after the initial removal of a line (modeling a fault triggered by some external event such as lightning strikes or sagging during summer), the subsequent failures are due to the dynamical evolution of the system variables in response to the removal of the faulty line(s).

These considerations motivate the problem of introducing in the power system (\ref{eq:swingequations}) a control law that, after the initial fault, can maintain the flows below the line capacities, such that failures in other lines of the power grid can be avoided. Here, we remark that, motivated by what occurs in the real system, we take into account a specific scenario, where the power grid starts from an equilibrium condition (i.e., the units are operating in a synchronous way) and an external event occurs to damage a line of the network. In response to such event, the network is driven towards another equilibrium that, according to the notion of dynamically-induced cascading failure \cite{schafer2018dynamically}, is also assumed to yield flows below the maximum capacity.

Let us, then, consider the following model for the controlled system:
\begin{equation}
\label{eq:swingequationscontrolled}
\begin{array}{l}
\frac{d\theta_i}{dt}=\omega_i\\
I_i\frac{d \omega_i}{dt}=P_i-\gamma_i\omega_i+\sum\limits_{(i,j)\in \mathcal{E}} K_{ij}\sin(\theta_j-\theta_i)+u_i
\end{array}
\end{equation}
\noindent where $u_i$ ($i=1,\ldots,N$) is the control input at each node, such that we can formulate our control problem as that of designing a control law $u_i$ guaranteeing that the flows obtained in response to the removal of a link of the network
do not meet the overload condition~(\ref{eq:overloadcondition})
for any $t$. Let $\bar{\mathcal{E}}$ indicate the original set of links of the power grid, $(i',j')$ the link subject to the initial fault, and $\mathcal{E}'$ the set of all the other network links, i.e., $\mathcal{E}'=\bar{\mathcal{E}}/(i',j')$, we require that system (\ref{eq:swingequationscontrolled}) with $\mathcal{E}=\mathcal{E}'$, starting from initial condition $\omega_i(0)=0$ and $\theta_i(0)=\theta_i^*$ generates flows $F_{ij}(t)$ that do not meet the overload condition (\ref{eq:overloadcondition}) for any $t$.


Summarizing, given a threshold $\alpha$, our control objective is to
design the controllers $u_i(t)$ in Eq.~(\ref{eq:swingequationscontrolled}) to prevent cascading failures, ensuring that, in the case of a fault in one of the power lines that drives the network out of synchrony for a period of time, Eq.~(\ref{eq:overloadcondition}) never occurs.

In the next section we offer a solution to the problem based on distributed controllers.

\section{Proposed control law}
\label{sec:proposedcontrollaw}

Let us assume that the power grid experiences a fault in the link $(i',j')$, which is then removed from the system. The removal of this link may trigger several additional line failures due to the propagation of overloads, as we now discuss. Consider the equilibrium $\mathbf{\theta}^*$ that corresponds to the solution of (\ref{eq:DCequilibrium}) with $\mathcal{E}=\bar{\mathcal{E}}$. This represents the equilibrium of the power grid, before the fault, and thus also the initial condition of the system, assuming that the system settled at it. However, after the removal of link $(i',j')$, the system is no longer at the equilibrium $\mathbf{\theta}^*$ and moves towards a new equilibrium, labeled as $\mathbf{\theta}^{**}$, that is given by Eq. (\ref{eq:DCequilibrium}) with $\mathcal{E}=\bar{\mathcal{E}}/(i',j')$. 

Suppose now that, at the new equilibrium $\mathbf{\theta}^{**}$, the steady-state values of the flows do not overcome the link capacities. However, the trajectory followed by the system may be such that the condition $|F_{ij}(t)|<C_{ij}$ is violated at some time $t$ for some $i$ and $j$, thus triggering the removal of the link and the succeeding cascading failure of the other lines.

Now notice that, during the transitory, after the initial fault at $(i',j')$ the power grid is out of synchrony. Since the flow $F_{ij}(t)$ in a generic link $(i,j)$ of the network depends on the difference between the angles and $\theta_j-\theta_i \approx (\omega_j-\omega_i)t$, then the larger is the difference between $\omega_j$ and $\omega_i$ the more likely the flow will experience quick fluctuations, eventually overcoming the threshold. For this reason, we propose a control strategy which reinforces synchrony in the network, thus limiting the rate of growth of the terms $\theta_j-\theta_i \approx (\omega_j-\omega_i)t$.

Specifically, we consider the following control law:
\begin{equation}
\label{eq:controldefinition}
u_{i}(t)=k_c\xi_i\sum\limits_{j=1}^N a_{ij}(\omega_j(t)-\omega_i(t))
\end{equation}

\noindent where the elements $a_{ij}$ define the topology of interactions among the controllers, such that they form a second layer, working in parallel with the physical one, where the energy exchanges among the power units take place. Here, we suppose that the topology of the control layer is the same of the physical wirings among the network nodes, and that if a link fails in the physical layer it also does in the control layer.
The parameters $\xi_i$ in (\ref{eq:controldefinition}) indicates in which nodes the control input is effectively applied. We consider two scenarios:
\begin{itemize}
\item \emph{full control}, where $\xi=1$ $\forall i=1,\ldots,N$;
\item \emph{pinning control}, where $\xi=1$ if $i\in \mathcal{N}_p \subset \mathcal{N}$, and $\xi=0$ otherwise.
\end{itemize}
In the first case, the control is applied to all units of the power grid. In the second case control is applied only to a subset of the network nodes, that we have indicated as $\mathcal{N}_p$. In analogy with the pinning control technique for network synchronization \cite{li2004pinning,chen2007pinning,delellis2011pinning}, where the input is applied to a subset of network nodes, we also refer to this scenario as pinning control, making clear that, with respect to the classical approach, here we do not consider a reference unit providing the trajectory towards which the system is driven. On the contrary, here, the pinned nodes are subject to control inputs with exclusively depend on the state of their neighboring nodes.

Notice that, as at the equilibrium we have that $\omega_i=0$ $\forall i=1,\ldots,N$, in both scenarios the equilibrium point does not depend on the presence/absence of the control (which is zero at the steady-state) nor on the given value of the gain $k_c$. 

\subsection{Control law principle}
\label{sec:approximation}


To get some insights on the effects of the control law, let us analyse an extremely simplified case study. Here we take into account full control ($\xi_i=i$ $\forall i$) and make several assumptions: 1) a linear approximation for the dynamics of the power grid is considered; 2) all generators and loads have the same parameter values, i.e., $\gamma_i=\gamma$, $I_i=I$ (in particular, $I=1$); 3) all lines have the same susceptance, i.e., $K_{ij}=K$.

Considering the linear approximation $\sin(\theta_j-\theta_i)\approx \theta_j-\theta_i$ and the other assumptions mentioned above, and replacing the control terms $u_i$ in (\ref{eq:controldefinition}), system (\ref{eq:swingequationscontrolled}) can be rewritten as:
\begin{equation}
\label{eq:swingequationsapprox}
\begin{array}{lll}
\frac{d\theta_i}{dt} & = & \omega_i\\
I_i\frac{d \omega_i}{dt} & = & P_i-\gamma_i\omega_i+\sum\limits_{(i,j)\in \mathcal{E}} K_{ij}(\theta_j-\theta_i)\\
& & +k_c\sum\limits_{(i,j)\in \mathcal{E}}(\omega_j-\omega_i)
\end{array}
\end{equation}

\noindent with $\mathcal{E}=\bar{\mathcal{E}}/(i',j')$. For the sake of notation simplicity, let us indicate with $\mathrm{L}_f$ the network Laplacian matrix encoding the connectivity of the set $\mathcal{E}=\bar{\mathcal{E}}/(i',j')$, i.e., the set of links after the removal of the line subject to the initial fault. In addition, since the network Laplacian is positive semi-definite, let us order its eigenvalues $\lambda_i$ as follows: $0=\lambda_1 \leq \lambda_2 \leq \ldots \leq \lambda_N$.

Let us now introduce the stack vectors $\mathbf{x}=[\theta_1, \omega_1,\ldots,\theta_N, \omega_N]^T$  and $\mathbf{P}=[0,\ldots,0,P_1,\ldots, P_N]^T$, and rewrite Eq. (\ref{eq:swingequationsapprox}) in compact form as:
\begin{equation}
\label{eq:compactform}
\begin{array}{l}
\frac{d\mathbf{x}}{dt}=\mathrm{A}_c\mathbf{x}+\mathbf{P}
\end{array}
\end{equation}

\noindent where $\mathrm{A}_c=\left [ \begin{array}{cc} \mathrm{0}_N & \mathrm{I}_N \\ -k \mathrm{L}_f & -\gamma\mathrm{I}_N-k_c\mathrm{L}_f \end{array} \right ]$.

The eigenvalues of the matrix $\mathrm{A}_c$, here indicated as $\mu_1,\ldots,\mu_{2N}$, can be analytically calculated. In fact, let us consider the matrix $\mathrm{T}$ containing the eigenvectors of $\mathrm{L}_f$, such that $\mathrm{T}^{-1}\mathrm{L}_f\mathrm{T}=\mathrm{D}$ where $\mathrm{D}=\mathrm{diag}\{\mu_1,\ldots,\mu_{2N}\}$. Consider then the matrix  $(\mathrm{I}_2\otimes\mathrm{T})^{-1}\mathrm{A}_c(\mathrm{I}_2\otimes\mathrm{T})$ that is similar to $\mathrm{A}_c$ and so has the same eigenvalues. We have that:
\begin{equation}
\label{eq:Asimilar}
(\mathrm{I}_2\otimes\mathrm{T})^{-1}\mathrm{A}_c(\mathrm{I}_2\otimes\mathrm{T})=\left [ \begin{array}{cc} \mathrm{0}_N & \mathrm{I}_N \\ -k \mathrm{D} & -\gamma\mathrm{I}_N-k_c\mathrm{D} \end{array} \right ]
\end{equation}

From (\ref{eq:Asimilar}) we calculate the characteristic polynomial of $\mathrm{A}_c$, which is given by:
\begin{equation}
\label{eq:CharacPoly}
\det(\mu \mathrm{I}_{2N}-\mathrm{A}_c) = \prod \limits_{i=1}^{N} (\mu^2+(\gamma+k_c\lambda_i)\mu+k\lambda_i)
\end{equation}
\noindent where $\lambda_i$ with $i=1,\ldots,N$ are the eigenvalues of $\mathrm{L}_f$. 

From (\ref{eq:CharacPoly}), it follows that the 
eigenvalues of $\mathrm{A}_c$ are given by:
\begin{equation}
\label{eq:eigenvaluesA}
\mu_{2i-1,2i}=\frac{1}{2}\left(-\gamma-k_c\lambda_i\pm\sqrt{(\gamma+k_c\lambda_i)^2-4k\lambda_i}\right)
\end{equation}

From the expression of the eigenvalues of the matrix $\mathrm{A}_c$, one can derive that the control gain $k_c$ can be selected such that the system modes are overdamped. 
In fact, the eigenvalues of $\mathrm{A}_c$ are given by Eq. (\ref{eq:eigenvaluesA}) which is obtained by solving:
\begin{equation}
\prod \limits_{i=1}^{N} (\mu^2+(\gamma+k_c\lambda_i)\mu+k\lambda_i)=0
\end{equation}

\noindent or, equivalently:
\begin{equation}
\prod \limits_{i=1}^{N} (\mu^2+2\xi_i\omega_i\mu+\omega_i^2)=0
\end{equation}

\noindent where $\omega_i^2=k\lambda_i$ and $\xi_i=\frac{\gamma+k_c\lambda_i}{2\sqrt{k\lambda_i}}$. Notice that, for $i=1$, as $\lambda_1=0$, we find $\mu(\mu+\gamma)=0$, which has real solutions. The system modes are overdamped if the damping coefficient $\xi_i$ is set $\xi_i>1$, $\forall i=2,\ldots,N$, which can be done selecting $k_c>\bar{k}_c$ with $\bar{k}_c$ given by:
\begin{equation}
\label{eq:kcsoglia}
\bar{k}_c=2\sqrt{\frac{k}{\lambda_2}}-\frac{\gamma}{\lambda_2}
\end{equation}
\noindent where $\lambda_2$ is the first non-zero eigenvalue of $\mathrm{L}_f$. Here, without loss of generality, we are implicitly assuming that the network associated to the Laplacian $\mathrm{L}_f$ is connected and so $\lambda_2 \neq 0$; otherwise, the initial fault would generate a set of nodes isolated from the rest of the network, yielding a scenario where synchronization cannot be restored using for the control layer the same topology of the power grid. Note that, since $\lambda_2$ depends on the location of the initial fault, and so does $\bar{k}_c$.

This analysis shows that increasing the control gain ${k}_c$ increases the damping in the system. The control action is, therefore, introducing in the power grid an energy dissipation term that makes possible to find a value of $k_c$ for which the trajectory going from one equilibrium point (corresponding to the situation before the occurrence of the fault) to the new equilibrium point (after the fault) does not met the overflow condition (recall that both equilibrium points are assumed to have flows below the maximum capacity). This is a broad principle that we expect to be effective also for the nonlinear original system, for which we resort to numerical simulations showing the suitability of the control action also in the more general case.

\section{Control of cascading failures in a small-size network}
\label{sec:controlsmallsize}

\begin{figure*}
\centering
\subfigure[]{\includegraphics[width=.32\textwidth]{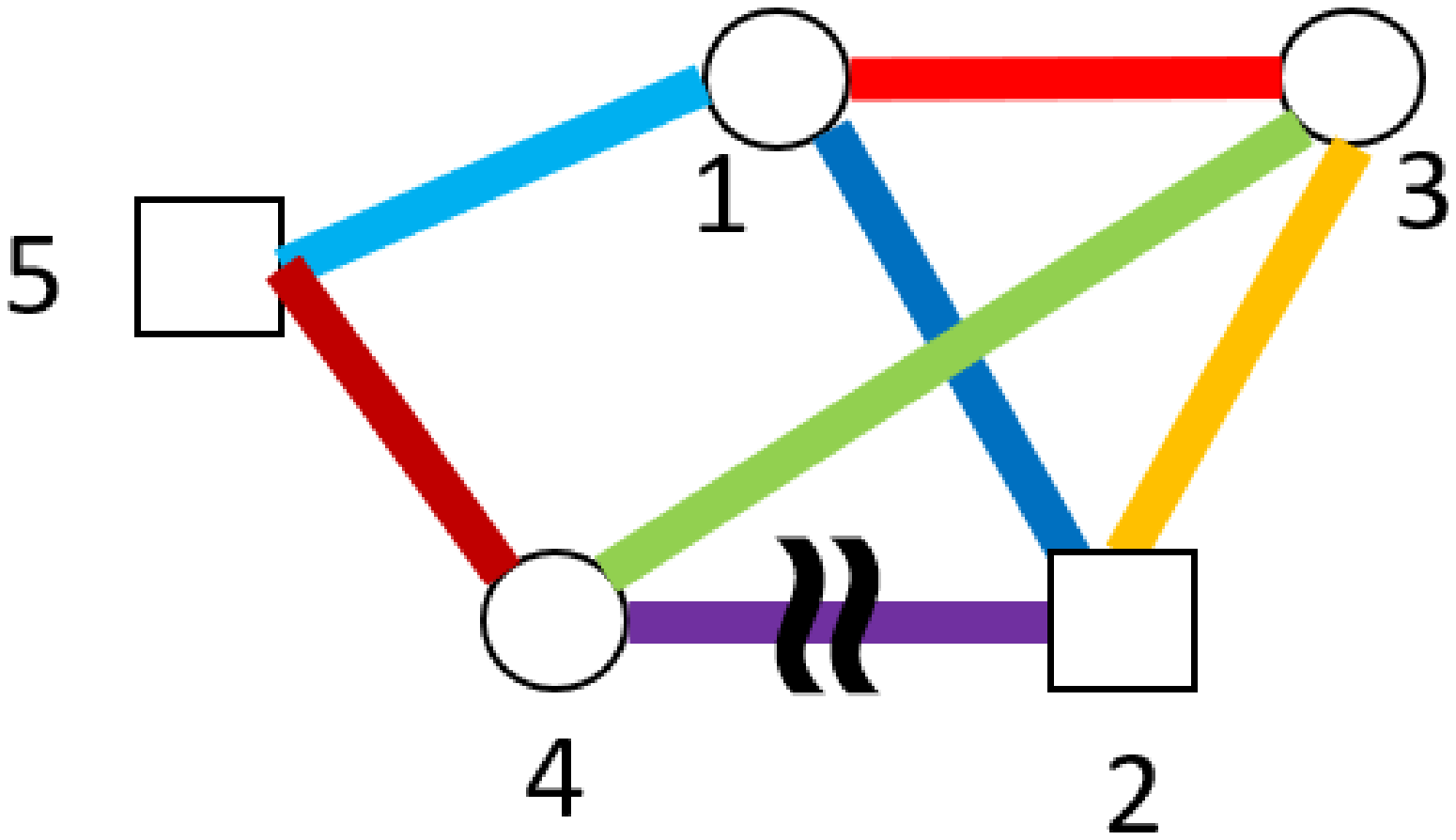}}
\subfigure[]{\includegraphics[width=.32\textwidth]{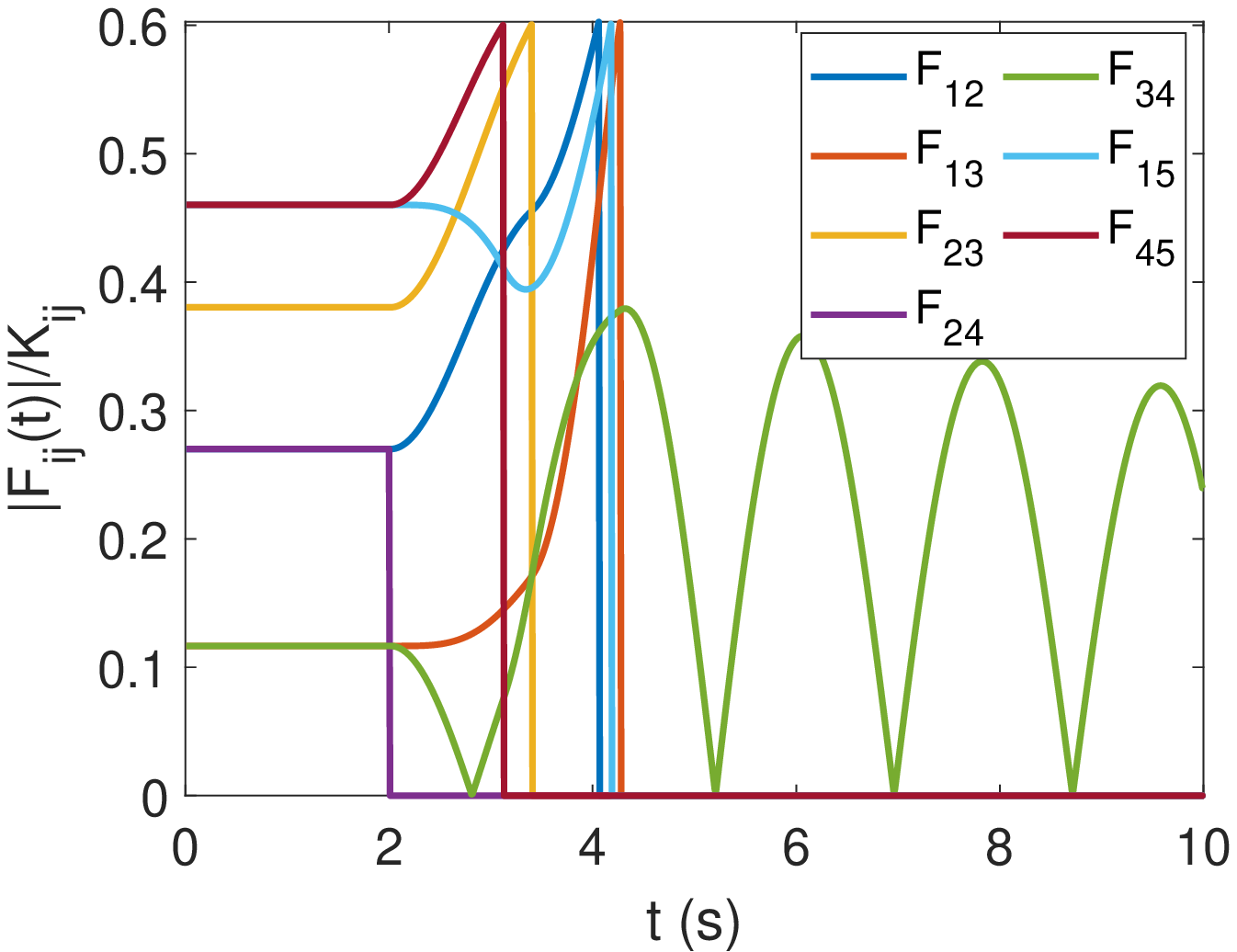}}
\subfigure[]{\includegraphics[width=.32\textwidth]{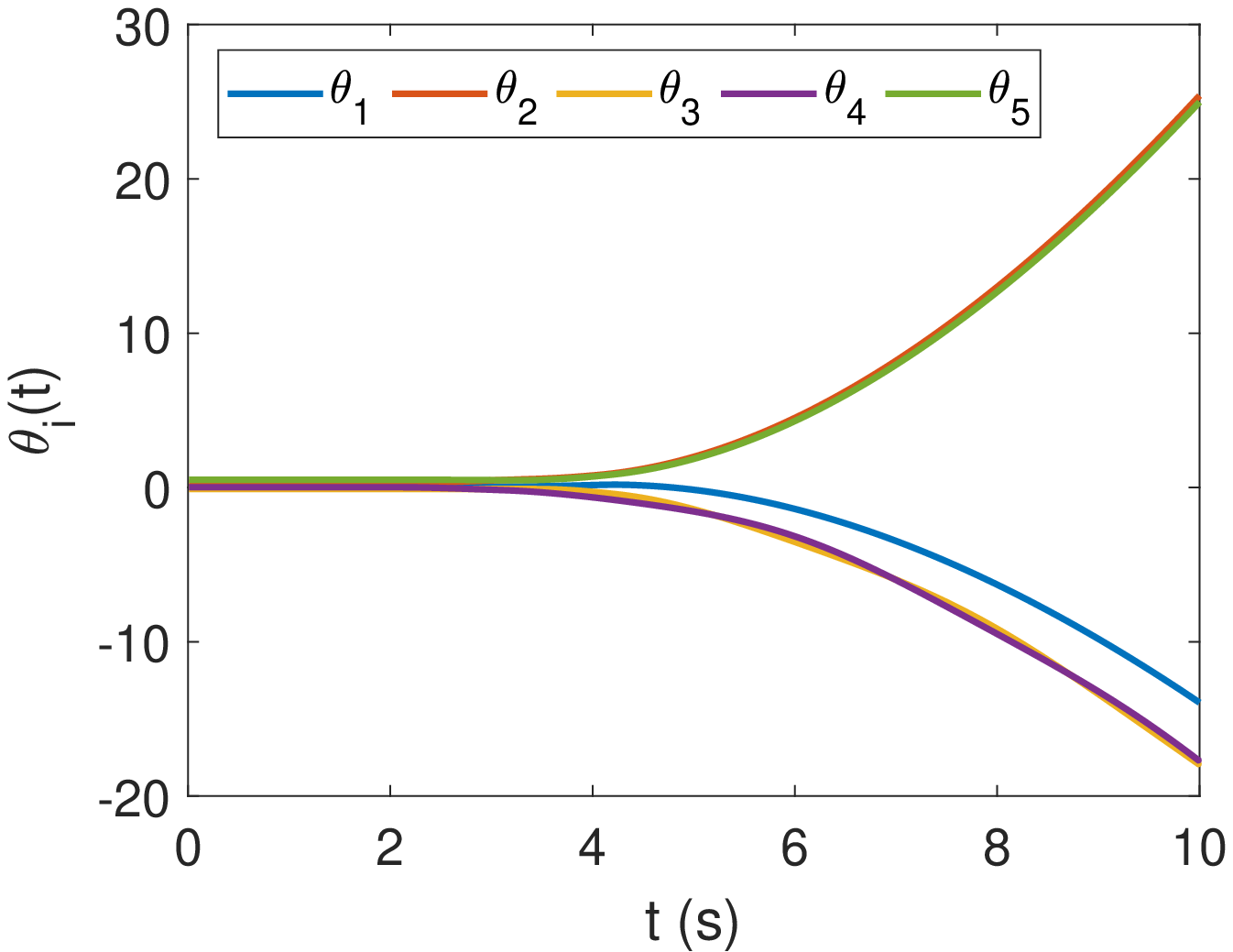}}
\subfigure[]{\includegraphics[width=.32\textwidth]{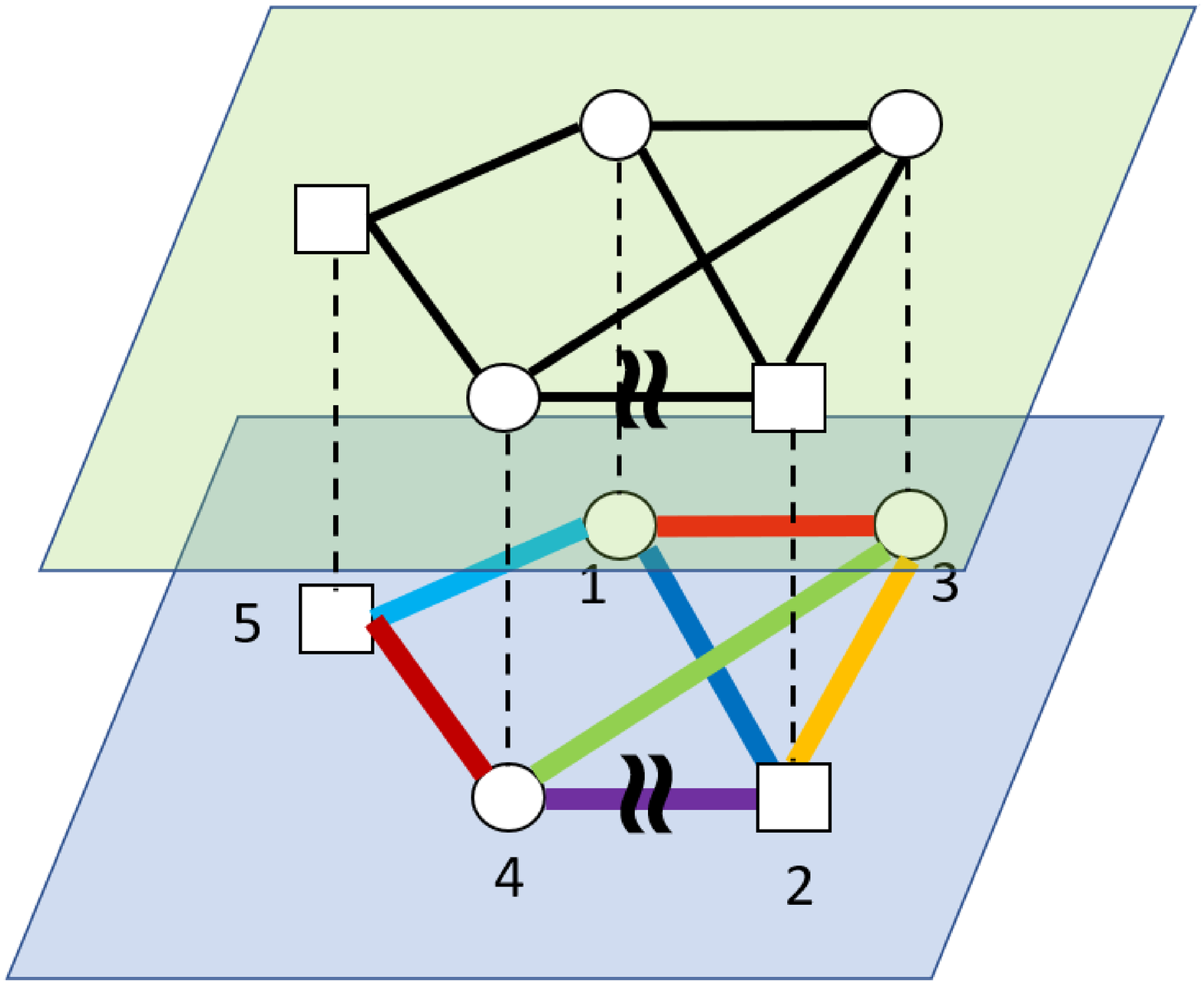}}
\subfigure[]{\includegraphics[width=.32\textwidth]{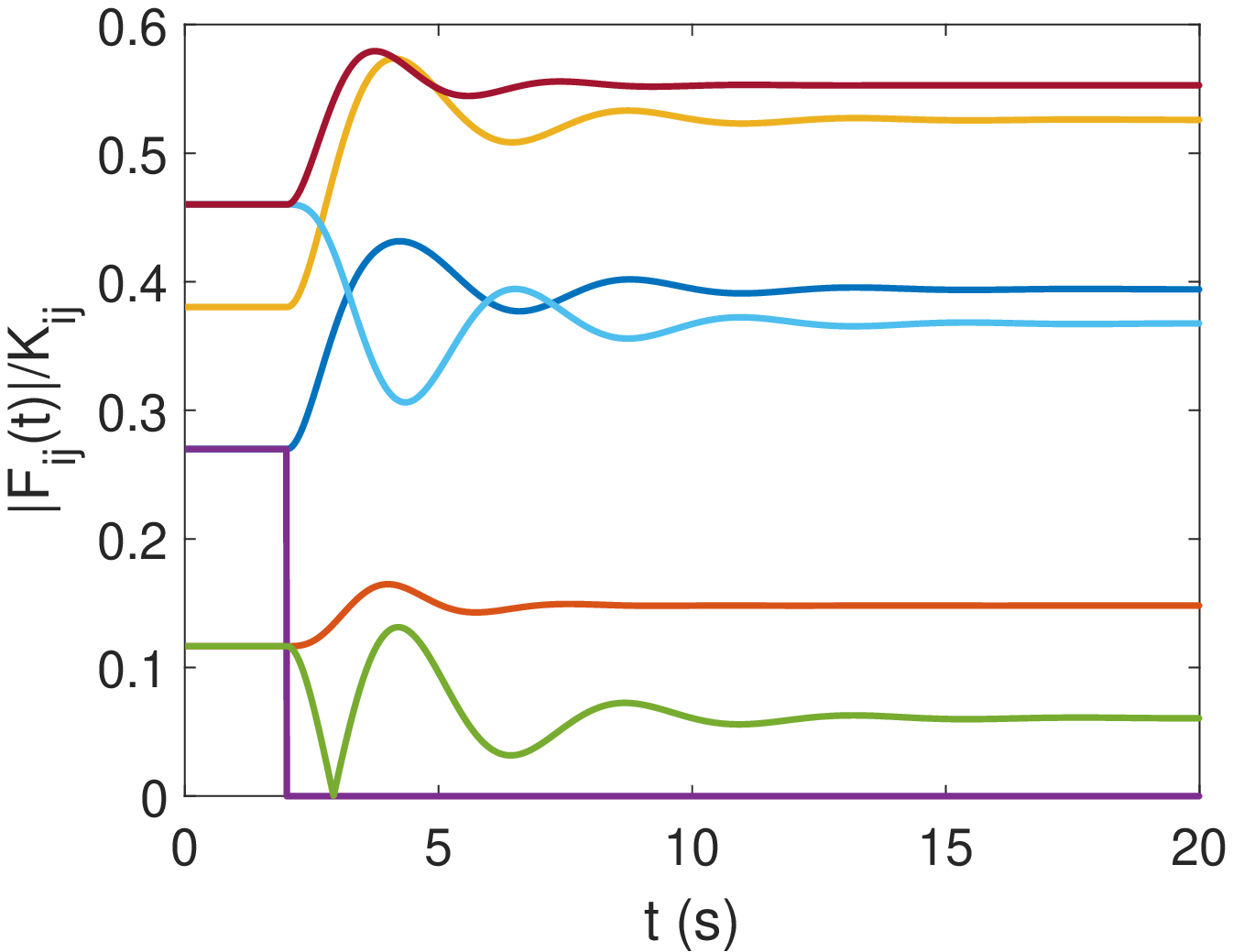}}
\subfigure[]{\includegraphics[width=.32\textwidth
]{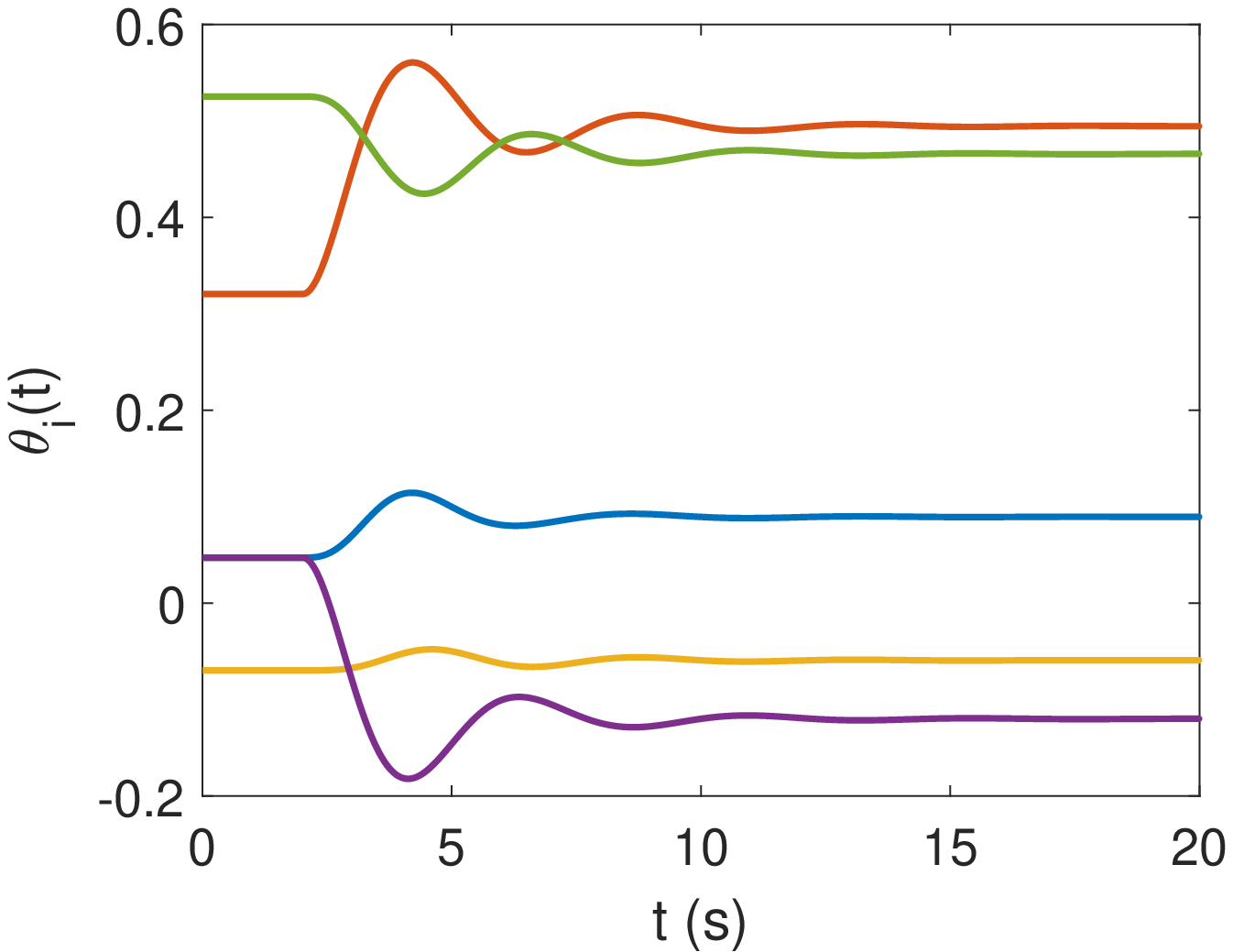}}
\subfigure[]{\includegraphics[width=.32\textwidth]{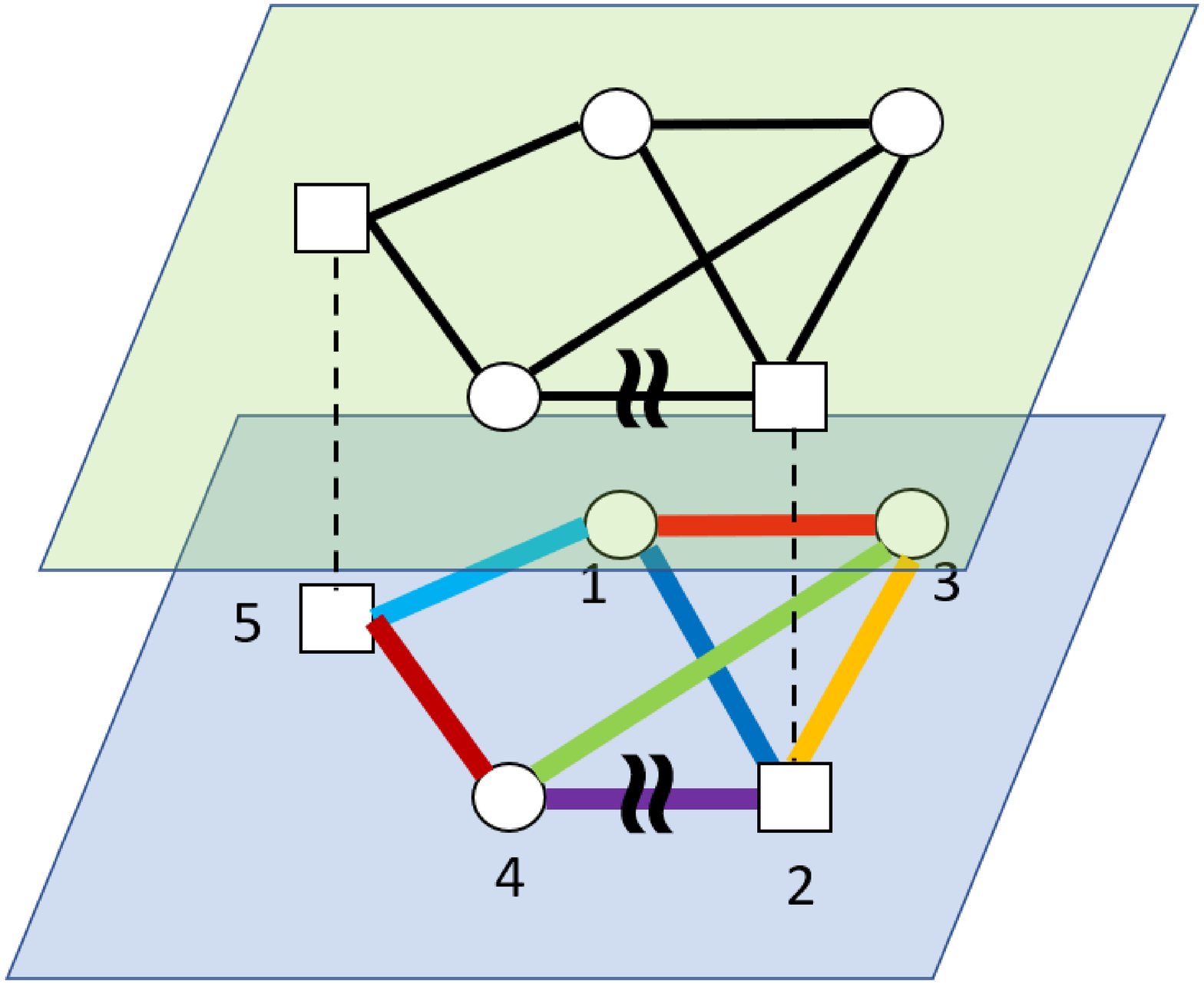}}
\subfigure[]{\includegraphics[width=.32\textwidth]{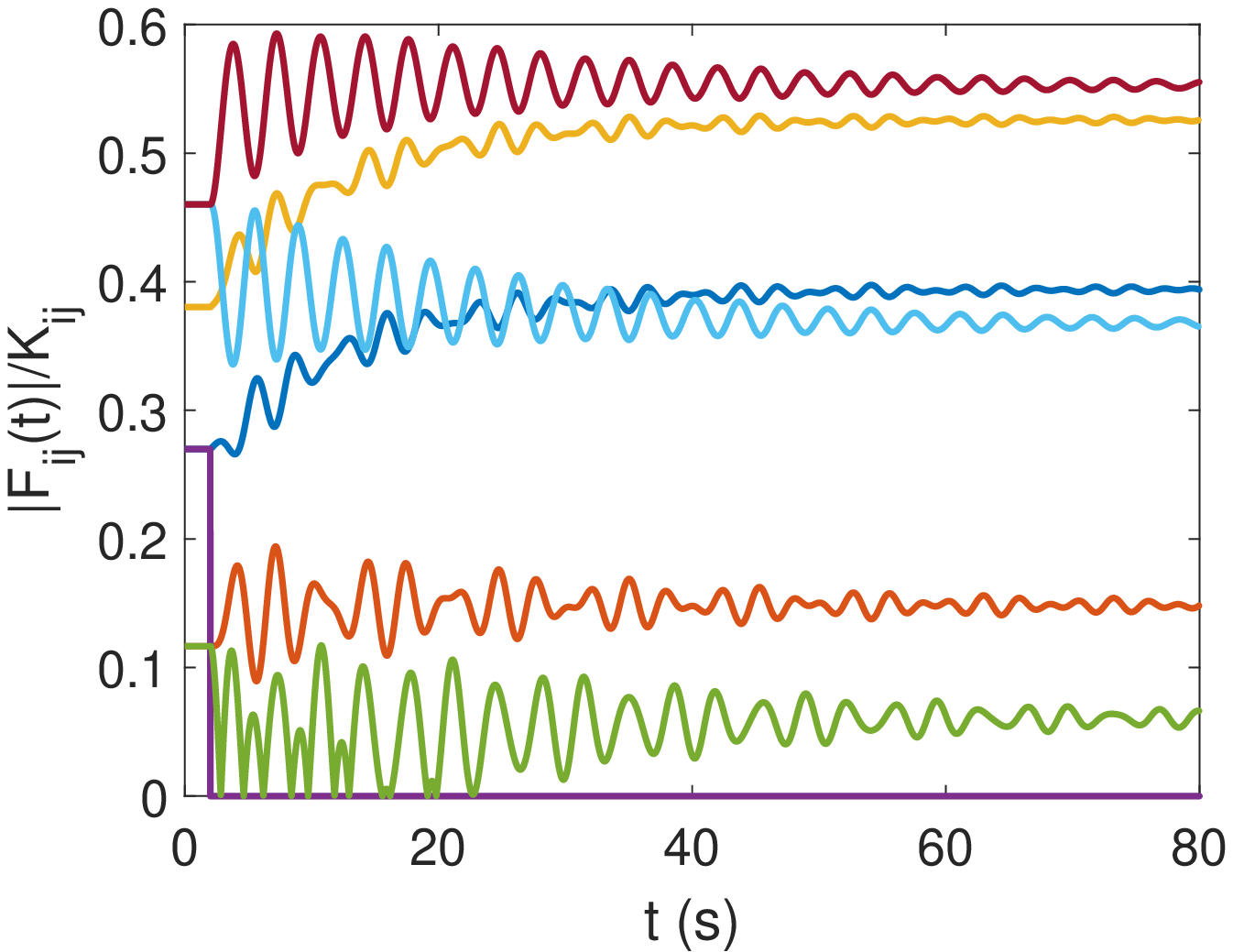}}
\subfigure[]{\includegraphics[width=.32\textwidth]{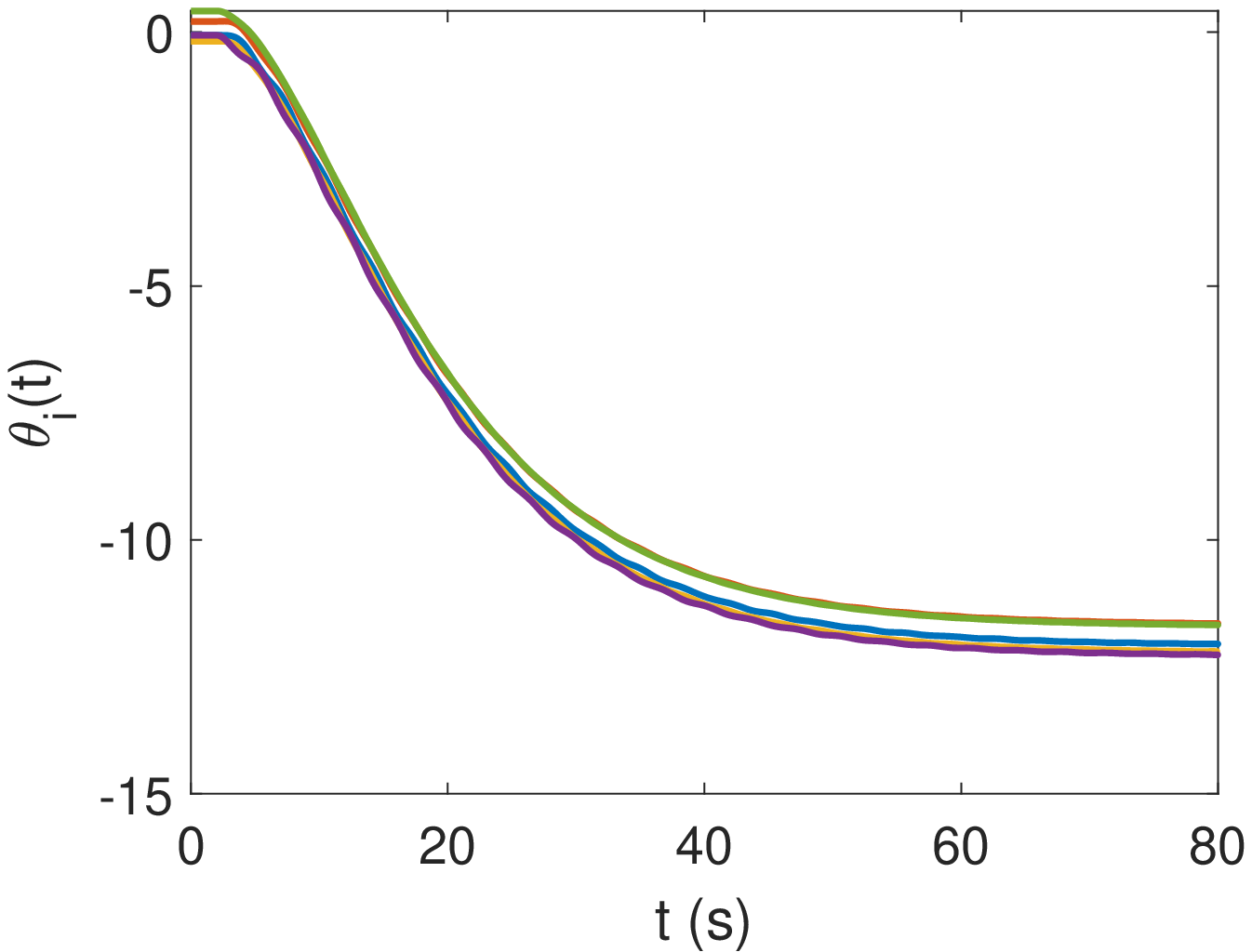}}
\caption{\label{fig:toymodel} Illustrative example of control of cascading failures in a five-node power grid. Scheme of the network without control (a), with full control (d), and with pinning control (g). Evolution of the flows without control (b), with full control (e), and with pinning control (h). Evolution of the angles without control (c), with full control (f), and with pinning control (i). The control gain was set to $k_c=0.5$ for full control and to $k_c=20$ for pinning control. The other parameters of the power system have been fixed as in \cite{schafer2018dynamically}: $\alpha=0.6$, $I_i=1$ $\forall i$, $gamma_i=0.1$ $\forall i$, $k=1.63$;
$P_i=-1$ for $i=\{1,3,4\}$ (loads), and $P_i=1.5$ for $i=\{2,5\}$ (generators).}
\end{figure*}

To illustrate the results of our control strategy, we first focus on a power grid of $N=5$ units, where it is possible to inspect the evolution of all state variables.  In particular, we refer to the example network discussed in \cite{schafer2018dynamically}, made of two generators and three loads (Fig. \ref{fig:toymodel}(a)). The network is indirected, so $K_{ij}=K_{ji}$. In addition, all weights are considered equal, such that $K_{ij}=ka_{ij}$, where $k$ represents the coupling strength and $a_{ij}$ are the elements of the adjacency matrix describing the network topology.

We start considering the initial fault in a given link of the network and discussing the behavior in absence of control. An initial fault is supposed to occur at time $t=2s$, after that the link $(2,4)$ is removed from the system, triggering several additional line failures due to the propagation of overloads (Fig.~\ref{fig:toymodel}(b)). We observe that the system starts from initial conditions given by the equilibrium point $\mathbf{\theta}^*$, corresponding to the solution of (\ref{eq:DCequilibrium}) with $\mathcal{E}=\bar{\mathcal{E}}$. After the removal of link $(2,4)$, the system is no more at equilibrium and moves towards a new equilibrium $\mathbf{\theta}^{**}$, given by Eq.~(\ref{eq:DCequilibrium}) with $\mathcal{E}=\bar{\mathcal{E}}/(2,4)$. Although at the new equilibrium $\mathbf{\theta}^{**}$, the steady-state flows do not overcome the line capacities, the trajectory followed by the system violates the condition $|F_{ij}(t)|<C_{ij}$ at the line $(4,5)$, triggering the removal of the line and a cascading failure involving four other edges.
The analysis of the time evolution of the angles $\theta_i(t)$ (Fig.~\ref{fig:toymodel}(c)) shows that during the transitory the power grid is effectively out of synchrony. This suggests that the control strategy (\ref{eq:controldefinition}), which reinforces synchrony in the network, thus limiting the rate of growth of the terms $\theta_j-\theta_i \approx (\omega_j-\omega_i)t$, may constitute an effective approach.

Let us now consider the controlled system (\ref{eq:swingequationscontrolled}), where the control layer works in parallel with the physical one and all nodes are controlled (full control, Fig.~\ref{fig:toymodel}(d)). The flows obtained when the control gain is set to $k_c=0.5$ are shown in Fig.~\ref{fig:toymodel}(e): they remain below the threshold value such that the system reaches the new equilibrium without triggering other failures. The angles $\theta_i(t)$ (Fig.~\ref{fig:toymodel}(f)) after a transitory reach the new steady-state equilibrium, thus recovering a synchronous state.

Consider now the pinning control strategy and, in particular, the case where the control input is applied only to generators (Fig.~\ref{fig:toymodel}(g)). As shown in Fig.~\ref{fig:toymodel}(h), illustrating the time evolution of the flows, also in this case, the control action is able to prevent the propagation of failures to other lines of the network. However, in this case, fulfilling the control goal requires a higher control gain $k_c$ ($k_c=20$ in the simulations illustrated in Fig.~\ref{fig:toymodel}(h)-(i)) and a longer transitory before the system settles in the new equilibrium point.

We now systematically analyse the network behavior for each possible location of the initial fault. The five-node power grid has seven links: two of them, namely $(1,3)$ and $(3,4)$, do not generate any succeeding failure; the lines $(1,5)$ and $(4,5)$, instead, lead to a static failure (i.e., at the equilibrium $\theta^{**}$ some of the flows meets the overload condition); and finally the three lines $(1,2)$, $(2,3)$ and $(2,4)$ produce a dynamically-induced cascading failure. 
Removal of any of the three links $(1,2)$, $(2,3)$ and $(2,4)$ generates a network that is still connected, such that, according to the analysis based on the linear approximation of Sec.~\ref{sec:approximation}, we expect to be able to control cascading failures for these links. This is confirmed by the inspection of the behavior of the controlled system (\ref{eq:swingequationscontrolled}) at different values of $k_c$. In particular, let us indicate with $n_c$ the number of lines failing after an initial fault in the link $(i',j')$. Since this number is also function of the control gain $k_c$, effectively one has $n_c=n_c(i',j',k_c)$.

\begin{figure}
\centering
\subfigure[]{\includegraphics[width=.44\textwidth]{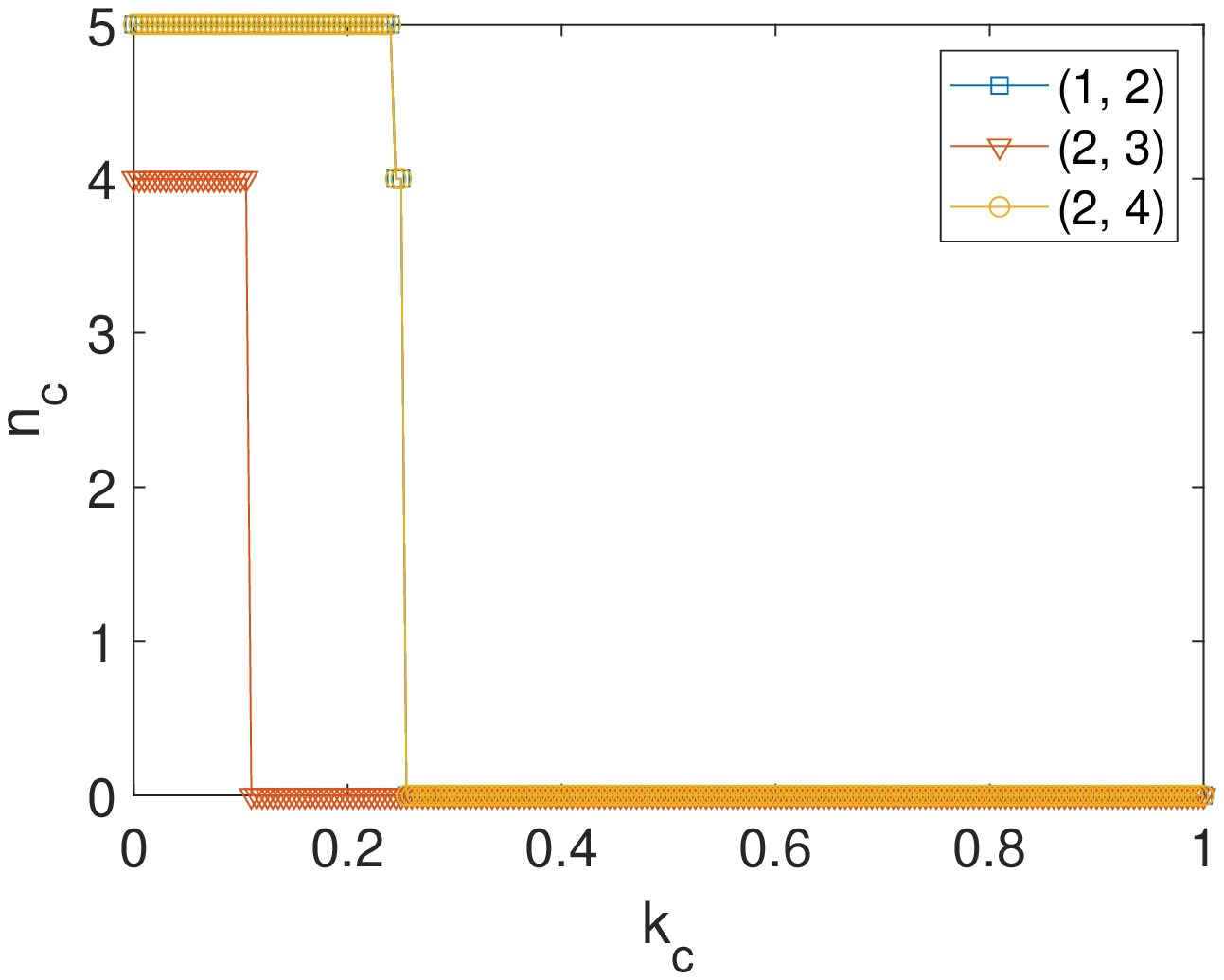}}
\subfigure[]{\includegraphics[width=.44\textwidth]{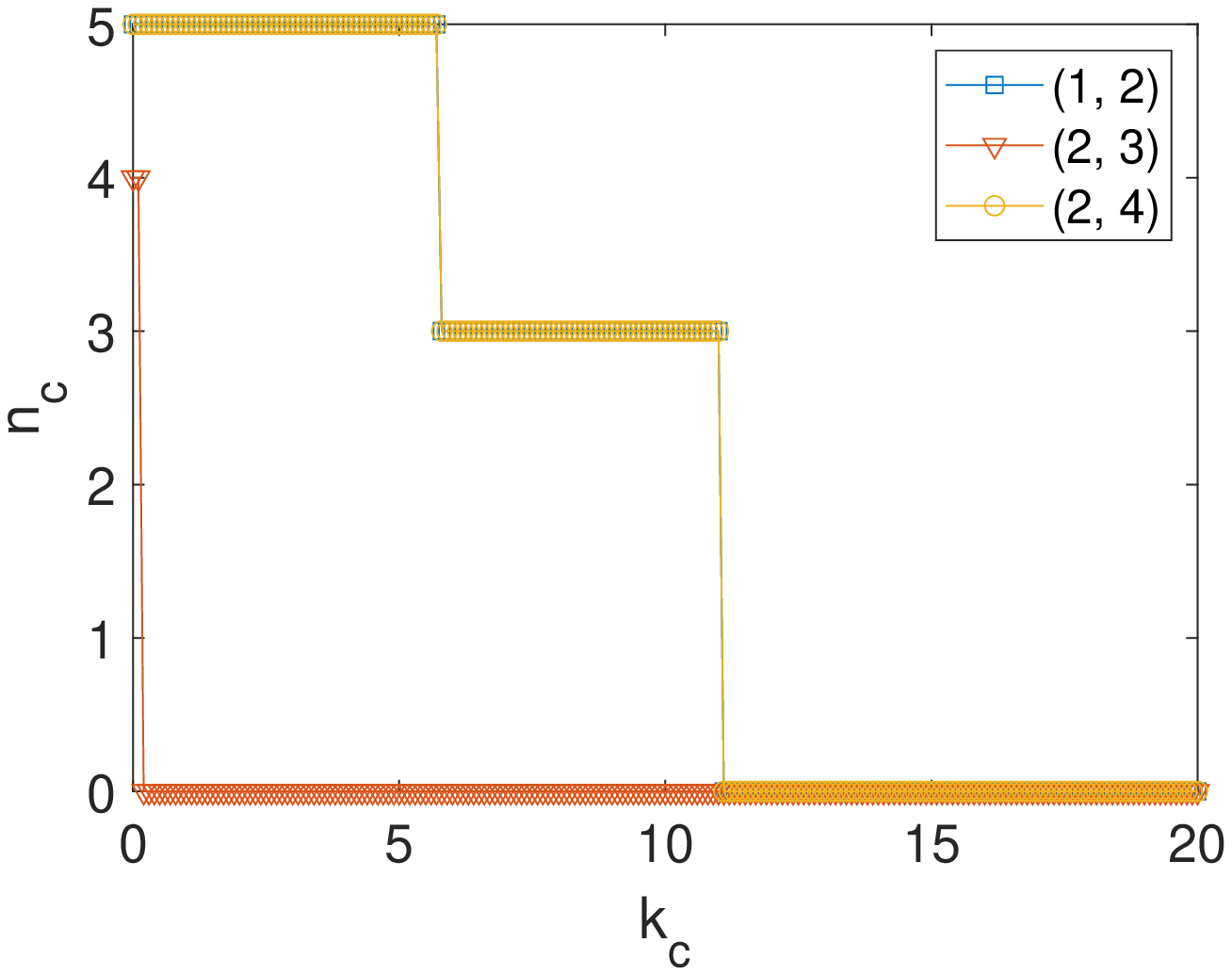}}
\caption{\label{fig:toymodel2} Number of failed lines, $n_c$, as a function of the control gain $k_c$ for an initial fault located in three different lines of the five-node power grid of Fig.~\ref{fig:toymodel}: (a) full control; (b) pinning control.}
\end{figure}

The parameter $n_c$ computed for full (Fig.~\ref{fig:toymodel2}(a)) and pinning (Fig.~\ref{fig:toymodel2}(b)) control demonstrates that, as the gain is increased, for any possible location of the initial fault, the control is able to avoid the failure of any other line of the power grid. The pinning case, generally, requires a higher gain as the control may act on a smaller set of nodes.

In addition, for the full control case, we compared the value of $k_c$ for which, in the numerical simulations of Fig.~\ref{fig:toymodel2}(a), $n_c$ vanishes with $\bar{k}_c$ obtained by Eq.~(\ref{eq:kcsoglia}).
For $(i',j')=(1,2)$ or $(i',j')=(2,4)$ Eq.~(\ref{eq:kcsoglia}) gives $\bar{k_c}=2.0997$, whereas, for $(i',j')=(2,3)$, $\bar{k_c}=1.7555$.
Although the computation of $\bar{k}_c$ is based on a linear approximation, we notice that it correctly predicts which lines require a higher control gain.


\section{Control of cascading failures in the Italian high-voltage (380kV) power grid}
\label{sec:ItalianGrid}

As a larger-size case study, we take into account the Italian high-voltage (380kV) power grid \cite{GENI}, which consists of $N=127$ nodes (34 generators and 93 loads) and $L=171$ links. The network is assumed to be undirected and unweighted, that is, $K_{ij}=K_{ji}=ka_{ij}=ka_{ji}$.
Similar assumptions have been considered in other works, e.g., \cite{rosato2007topological,filatrella2008analysis,fortuna2012network,tumash2019stability,totz2020control}. In our simulations, the  damping coefficient $\gamma$ and the overload threshold parameter $\alpha$ have been set to $\gamma=0.1$ and $\alpha=0.6$, as in \cite{schafer2018dynamically}. For the load nodes, the power has been normalized to the value $P_i=-1$, and, consequently, for the generation nodes, the power has been set to $P_i=2.7353$ to guarantee that the network is balanced, i.e., $\sum\limits_{i=1}^N P_i=0$. Finally, the coupling strength has been set to $k=15$, such that, in the absence of faults, the network is synchronized. 

First, we have analyzed the network in absence of control, $u_i=0$, by systematically initializing a fault in each line of the power grid and determining static and dynamic failures. The static failures have been ascertained by calculating the equilibrium point of the power grid using (\ref{eq:DCequilibrium}), checking the overload condition (\ref{eq:overloadcondition}) at the obtained steady-state value, eventually removing each overloaded line, and iteratively repeating the procedure until no other line meets (\ref{eq:overloadcondition}). The dynamical failures have been obtained with a similar procedure, where, however, the overload condition (\ref{eq:overloadcondition}) is checked at any time $t$ using the result of the integration of Eqs. (\ref{eq:swingequations}) rather than the equilibrium point. This analysis shows that there are twenty-two lines that, when subject to an initial fault, lead to further failures in the dynamical model, among which six yield a cascading failures also in the static model. 

\begin{figure}
\centering
\includegraphics[width=.65\textwidth]{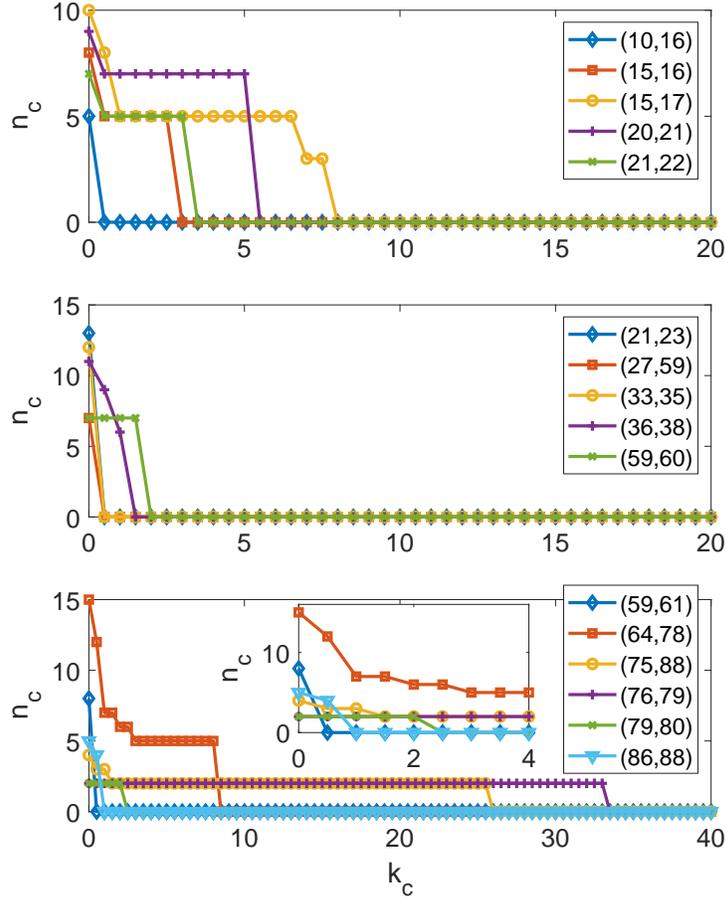}
\caption{\label{fig:itaPG} Number of failed lines, $n_c$, as a function of the control gain $k_c$ for an initial fault located in any of sixteen different lines of the Italian high-voltage (380kV) power grid producing a dynamically-induced cascade.}
\end{figure}

We have then applied the full control strategy to the sixteen lines where the cascading failures are only due to the dynamics of the flows and not to their steady-state value, while monitoring the number of failed lines $n_c=n_c(i',j',k_c)$ as a function of the location of the initial fault and $k_c$. Notice that, similarly to what observed for the five-node example, also in this case, after removal of any of these lines the network remains connected, such that the control law (\ref{eq:controldefinition}) may be applied.

The curve of $n_c$ with respect to $k_c$ is shown in Fig.~\ref{fig:itaPG}, which demonstrates that all dynamically-induced cascading failures can be controlled for a sufficiently high value of the control gain. Also for the Italian high-voltage power grid, we have computed $\bar{k}_c$ from (\ref{eq:kcsoglia}). The results, given in Table~\ref{tab:my_label}, show that, despite the strong approximation introduced, still the analysis provides a conservative value for the gain, in the sense that, for each of the sixteen lines producing a dynamically-induced cascading failure, setting $k_c>\bar{k}_c$ always prevents the fault to spread into other lines of the network. In addition, the most 'critical' lines, i.e., those requiring the highest values of $k_c$, are correctly individuated.

\begin{table}[]
    \centering
    \begin{tabular}{|c|c|c|c|}
         \hline 
         Link &  $\bar{k}_c$ &  Link &  $\bar{k}_c$\\
         \hline (10,16) & 55.1839 &  (15,16) &  55.1353 \\
         \hline (15,17) &  55.1312 &  (20,21) &   55.1414 \\
         \hline (21,22) &  55.1327 &  (21,23) &  55.2472  \\
         \hline (27,59) &  55.5946 &  (33,35) & 55.3000  \\
         \hline (36,38) & 55.2745 &  (59,60) & 55.2189  \\
         \hline (59,61) & 55.7898  &  (64,78) & 58.0795   \\
         \hline (75,88) & 59.8266  &  (76,79) & 62.3968   \\
         \hline (79,80) &  58.0187 &  (86,88) & 55.2515 \\
         \hline
    \end{tabular}
    \caption{Values of $\bar{k}_c$ for the Italian high-voltage (380kV) power grid.}
    \label{tab:my_label}
\end{table}

To conclude this section, we discuss the results on the application of the pinning control strategy to the Italian high-voltage power grid. Here, we have found that acting exclusively on the set of generators, that is, setting $\xi=1$ $\forall i \in \mathcal{N}_g$, is not sufficient to control all the lines subject to dynamically-induced failures.

\begin{figure}
\centering
\includegraphics[width=.65\textwidth]{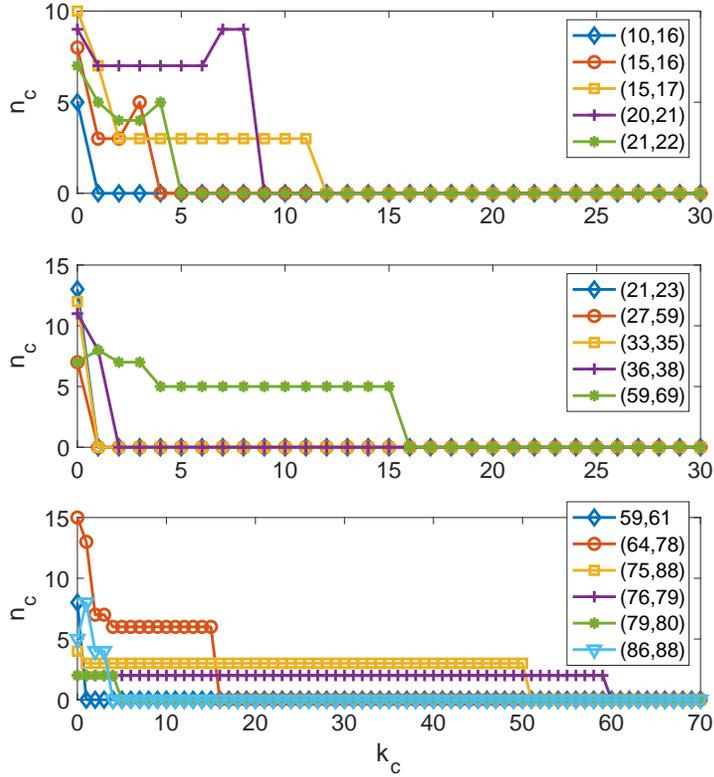}
\caption{\label{fig:pgItaPinning} Pinning control of the Italian high-voltage (380kV) power grid. Number of failed lines, $n_c$, as a function of the control gain $k_c$ for an initial fault located in any of the sixteen lines of the grid producing a dynamically-induced cascade.}
\end{figure}

We have then considered the set of pinned nodes as formed by that of generators plus further nodes selected among the remaining ones, i.e., the loads. An optimal selection of the pinned nodes aimed at minimize their number, although particularly interesting, would require an extensive search among all the possible configurations of pinned nodes that is beyond the scope of this work. For this reason, we have heuristically determined a set of pinned nodes, namely $\mathcal{N}_p=\mathcal{N}_g \cup \{15,16,20,21,64,75,76,79,80,86,88\}$, that,
without any guarantee of optimality, still prove to fulfill the control objective. In fact, for each of the sixteen lines yielding a cascading failure, using pinning control with this set of pinned nodes, the number of failed nodes $n_c$ approaches zero for sufficiently large $k_c$  (Fig.~\ref{fig:pgItaPinning}). As noticed for the five-node network, also in this case, we observe that pinning control, generally, requires higher values of the control gain. The lines requiring the highest values of $k_c$ are the same of the full control case.

\section{Control of cascading failures in the IEEE 118-bus test case}
\label{sec:IEEE118}

The last case study considered in this paper refers to the IEEE 118-bus system. This test case represents an approximate description of the American Electric Power system in the Midwestern US as of December 1962, with $N=118$ nodes (54 synchronous machines and 64 load stations) and 186 lines \cite{christie118}. Data are available for several parameters of the system, such as the inertia of the synchronous machines and the susceptances associated to the lines. This case study has not to be intended as a realistic simulation, as we are modeling the system with the swing equations (\ref{eq:swingequations}) which are an approximated model where also the loads are modeled as synchronous machines. On the contrary, it has to be viewed as a proof of concept that the control approach introduced in this work may be applied to networks that are weighted and have heterogeneous parameters governing the dynamics of the units.

\begin{figure}
\centering
\subfigure[]{\includegraphics[width=.44\textwidth]{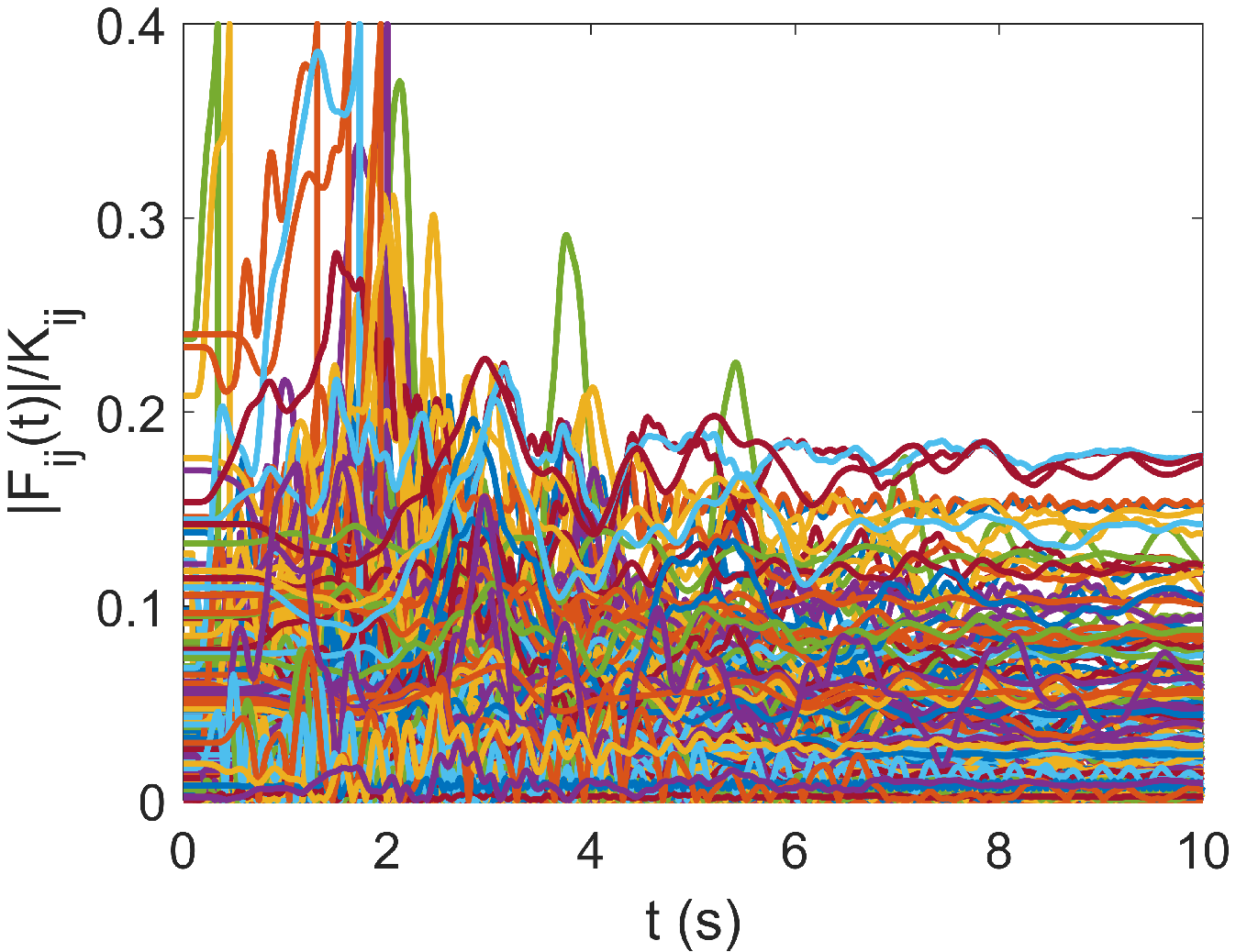}}
\subfigure[]{\includegraphics[width=.44\textwidth]{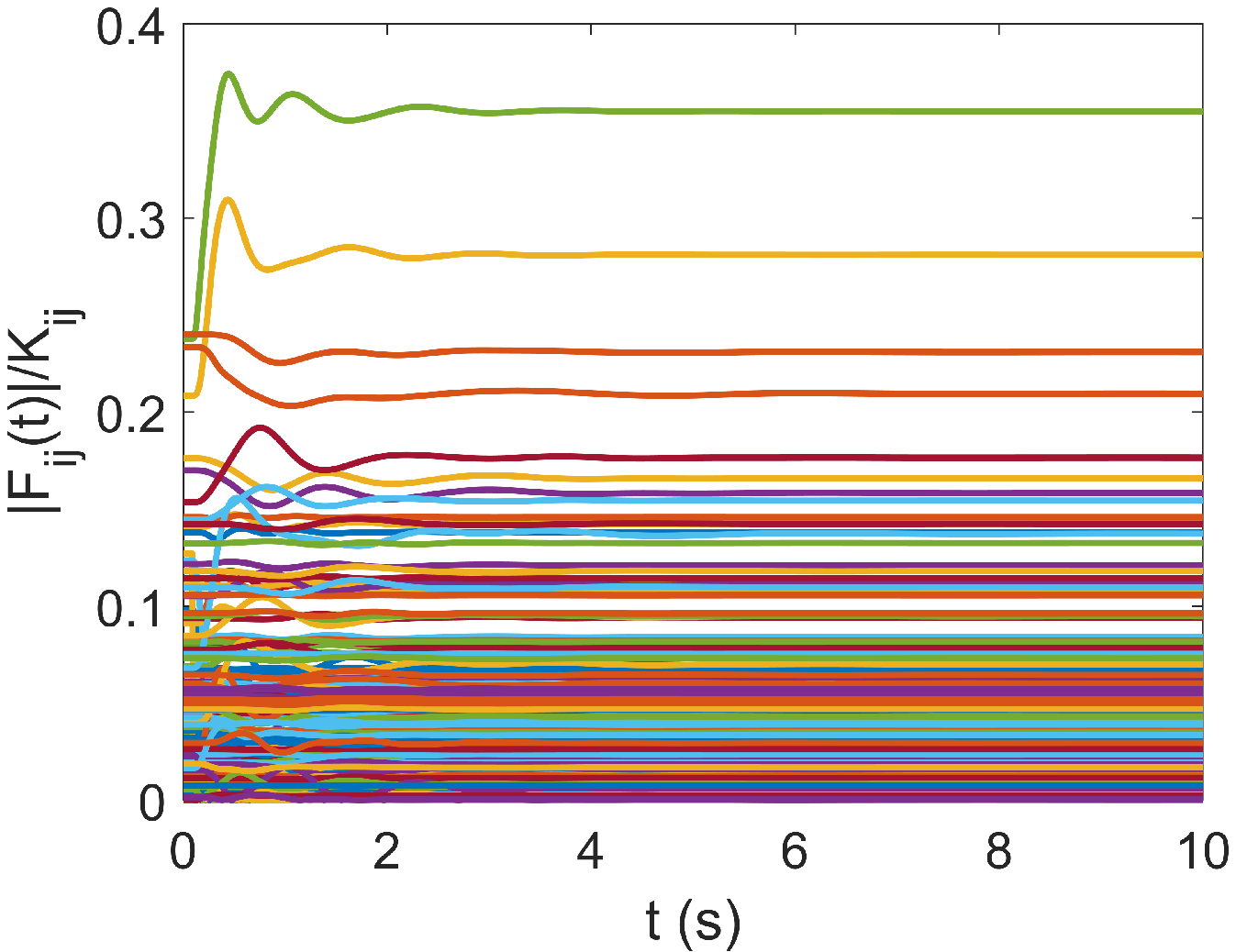}}
\subfigure[]{\includegraphics[width=.44\textwidth]{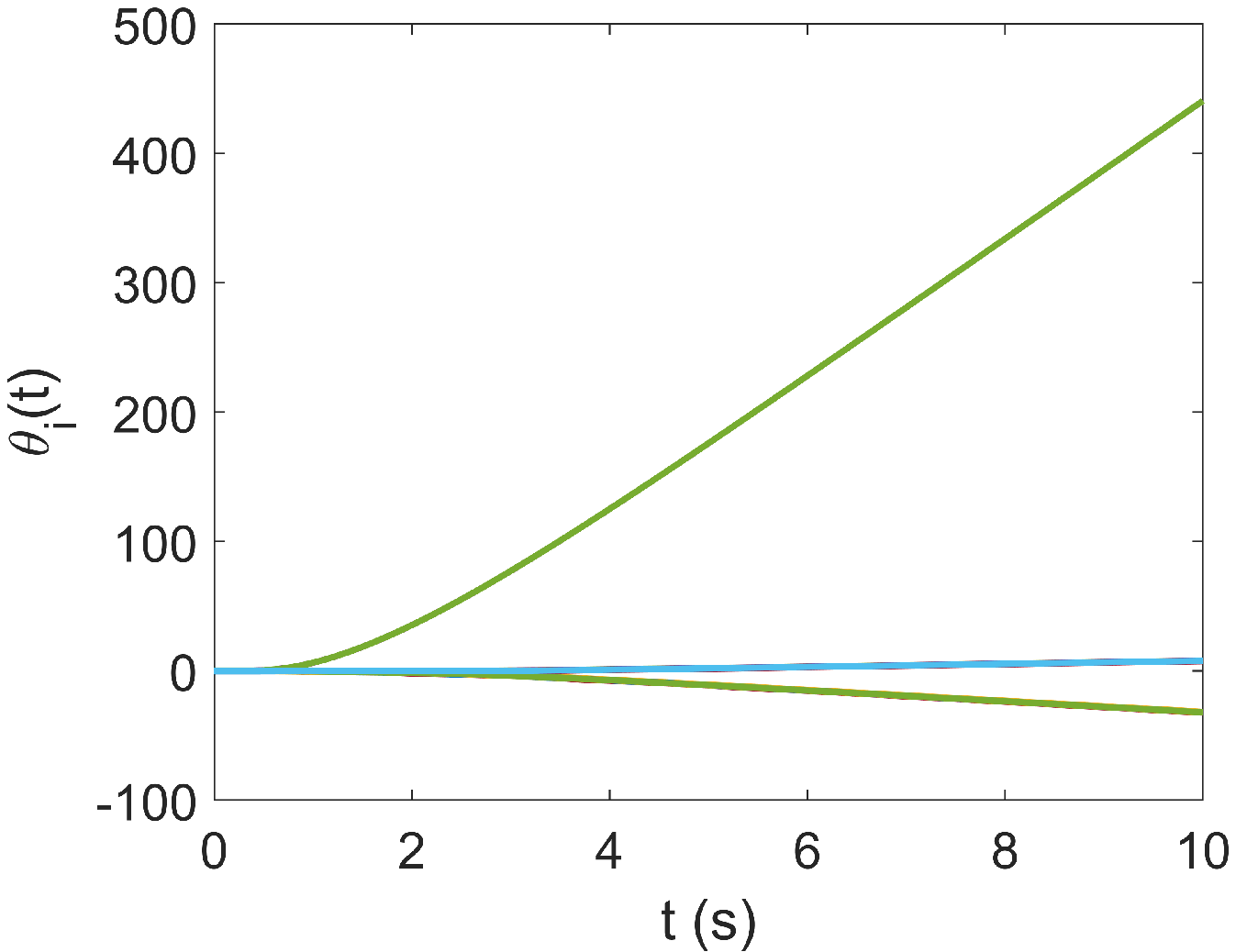}}
\subfigure[]{\includegraphics[width=.44\textwidth]{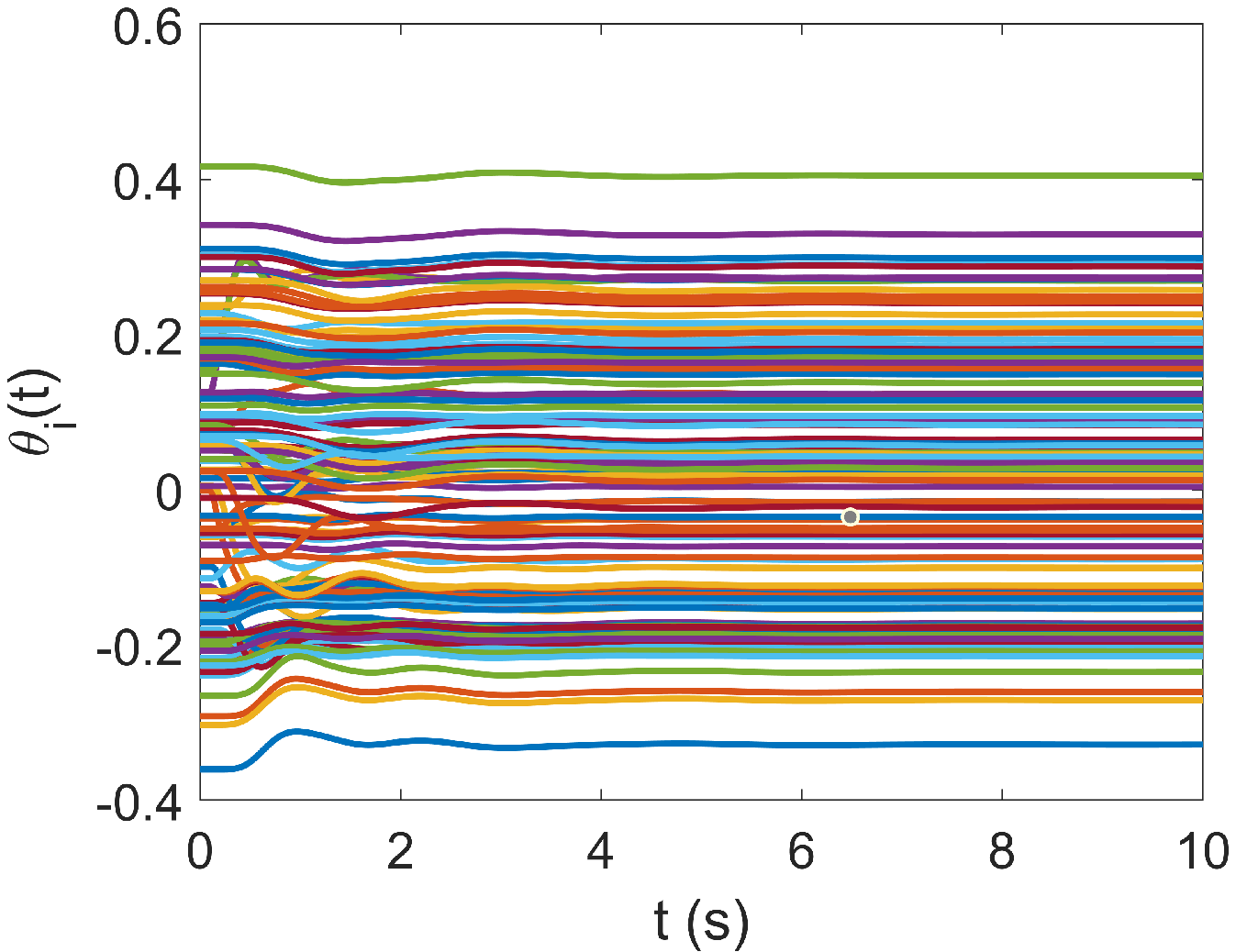}}
\caption{\label{fig:ieee118} Control of cascading failures in the IEEE 118-bus test case. Evolution of the flows without (a) and with control (b). Evolution of the angles without (c) and with control (d). The control gain was set to $k_c=0.5$. The initial fault is located in line 31 of the power grid.}
\end{figure}

In particular, we have considered a fault located in line 31 (that connects nodes 23 and 25) of the IEEE 118-bus test case and simulated the power grid setting the inertia of the 54 synchronous machines, the power $P_i$ for each node, and the line susceptances $K_{ij}=B_{ij}$ as in \cite{christie118}. For the remaining parameters, one has to take into account that the swing equations provide an approximated model of the grid with parameters that are not available in the data (e.g., the inertia value for nodes which are not synchronous machines, but are described as such in the model) or need re-adaptation because of the model assumptions. We have, thus, empirically selected the inertia of the remaining nodes as $I_i=I=0.064$, corresponding to the average value of that of the synchronous machines, and the damping coefficient as $\gamma_i=\gamma=0.05$ $\forall i$, such that in the absence of faults the power grid is synchronized. Finally, the overload threshold has been set to $\alpha=0.4$. In the absence of control, a fault in line 31 triggers the failures of seven other lines (Fig.~\ref{fig:ieee118}(a) and (c)). When the control law~(\ref{eq:controldefinition}) is applied to all nodes of the grid (full control), then all flows remain below the maximum capacity (Fig.~\ref{fig:ieee118}(b)) and the angles $\theta_i(t)$ converge to the new equilibrium point (Fig.~\ref{fig:ieee118}(d)).

\section{Conclusions}
\label{sec:Conclusions}

In this work we have investigated the problem of controlling dynamically-induced cascading failures in power grids, proposing a solution based on distributed controllers forming a second layer acting in parallel with the physical one. The suitability of the approach has been demonstrated through the analysis of a series of case studies including the Italian high-voltage power grid and the IEEE 118-bus system. 

In particular, we have focused on those failures which are induced solely by the dynamics of the power grid. For these failures, the steady-state distribution of the flows after the initial fault of a line of the network does not produce any overload in other lines. The problem of avoiding also static failures can be addressed by considering a different distribution of the powers at the network nodes or adopting a control law which also changes the system equilibrium point.

As the application of the control law to all network nodes may be considered a limitation of the approach, we have also considered the case where only a subset of the nodes are subject to a control term, and shown that, although generally requiring higher gains, also this strategy is viable. As the computation of the subset of nodes to control in order to prevent cascading failures is not straightforward, a possible direction for future work is the definition of automatic techniques to find the minimum number of such nodes.







\end{document}